\documentclass{article}
\usepackage[utf8]{inputenc}
\usepackage{amsmath}
\usepackage{amssymb}
\usepackage{amsfonts}
\usepackage{esvect}
\usepackage{cite}
\usepackage{physics}
\usepackage{mathrsfs}
\usepackage{authblk}
\usepackage{geometry}
\usepackage{abstract}
\usepackage{xcolor}
\usepackage{bbm}

\usepackage{graphicx}               % for figures
\usepackage{float}                  % to use the "H" placement for figures

\usepackage{csvsimple}
\usepackage{tabularx}
\usepackage{subcaption}
\usepackage{authblk}

\title{\bf Representations of quadratic Heisenberg-Weyl algebras and polynomials in the fourth Painlev\'e transcendent}
\author{\large Ian Marquette \footnote{i.marquette@latrobe.edu.au} }
\affil{ Department of Mathematical and Physical Sciences, La Trobe University, \\ Bendigo 3552, Victoria Australia  }

\usepackage[colorlinks=true,urlcolor=black]{hyperref}

\begin{document}

\maketitle
\begin{abstract}
We provide new insights into the solvability property of an Hamiltonian involving of the fourth Painlev\'e transcendent and its derivatives. This Hamiltonian is third order shape invariant and can also be interpreted within the context of second supersymmetric quantum mechanics. In addition, this Hamiltonian admits third order lowering and raising operators, which is a direct consequence of the shape invariance property. We will consider the case when this Hamiltonian is irreducible i.e. when no special solutions exist for given parameters $\alpha$ and $\beta$ of the fourth Painlev\'e transcendent $P_{IV}(x,\alpha,\beta)$. This means that the Hamiltonian does not admit a potential in terms of rational functions ( or hypergeometric type of special functions ) for those parameters. In such irreducible case, the ladder operators are as well involving the fourth Painlev\'e transcendent and its derivative. An important case for which this occurs is when the second parameter (i.e. $\beta$) of the fourth Painlev\'e transcendent $P_{IV}(x,\alpha,\beta)$ is strictly positive i.e. $\beta>0$. This Hamiltonian has been studied for all hierarchies of rational solutions that comes in three families connected to the generalised Hermite and Okamoto polynomials. Those occur when the parameter $\beta<0$ and specific values of $\alpha$ and $\beta$. The explicit form of ladder, the associated wavefunctions involving exceptional orthogonal polynomials and recurrence relations were also completed described. Other solutions in terms of hypergeometric type of special functions exist which have been discuss as well. Much less is known for case beyond those rational solutions and in particular in the irreducible case, in particular for the excited states. Here, we develop a description of the induced representations based on various commutator identities for highest and lowest weight type representations. We also provide for such representations new formula concerning the explicit form of the related excited states from point of view of the Schrodinger equation as two variables polynomials that involve the fourth Painlev\'e transcendent and its derivative. We present different formula of those polynomials in the case of highest weight and lowest weight type representations. To our knowledge this is first time such polynomials are considered for the excited states of the model.
\end{abstract}

\section{Introduction}

In context of quantum Hamiltonians various approaches to find the solution of their corresponding Schr\"odinger equation have been introduced. Some of those approaches rely on analytical and algebraic definitions of exact and quasi-exact solvability \cite{tur88, ush94,zha12}. They build on different ideas such as the existence of an underlying hidden algebra, Bethe Ansatz equations and invariant spaces of polynomials. Other approaches can be used to solve quantum mechanical systems which include Darboux-Crum and Krein-Adler types of transformations and their related intertwining relations, factorization relations and superalgebras \cite{coo00}. An important class of algebraic approaches concerns Liouville integrability and superintegrability where Abelian and non Abelian algebraic structures within the Lie theory or one of its generalization play the role of symmetry algebras \cite{mil13}. In this paper, we will consider the case of ladder operators, and in particular of third order ladder operators and its related Hamiltonian. In such setting, the Hamiltonian is  not described by an algebraic form but written in terms of the fourth Painlev\'e transcendents $P_{IV}(x,\alpha,\beta)$ \cite{ves93,and00, car04, ber14}. Some insight into zero modes was provided. This one dimensional Hamiltonian also appear in context of superintegrable systems as building block which makes it an important model \cite{mar09,mar10,mar11}. Further details on the zero modes of its related superpartner was also provided \cite{mar09}. 

The case where $\alpha$ and $\beta$ take values for which the fourth Painlev\'e transcendent admits rational solution was completely studied in recent papers \cite{mar16,hus21,zel22}. Those cases which relate to generalized Hermite and Okamoto polynomials occur for ($\alpha=2m+n$, $\beta=-2(n-\frac{1}{3})^2$), ($\alpha-m-2n$, $\beta=-2(m-\frac{1}{3})^2$) and ($\alpha=n-m$, $\beta=-2(m+n+\frac{1}{3})^2$). The connection with exceptional orthogonal polynomials was also highlighted. This paper will deal with irreducible cases i.e. when the fourth Painlev\'e transcendent does not admit reduction.

For a broader discussion of the problem of classifying systematically systems with ladder operators we refer the reader to \cite{nik97,mar19}. Deformed Heisenberg-Weyl algebras ( span$\{c,c^{\dagger},H,\mathbbm{1}\}$) take the generic form

\begin{subequations}
\begin{align}
 [H,c]=-a c \\
 [H,c^{\dagger}]=a c^{\dagger} \\
 [c,c^{\dagger}]=F(H) 
 \end{align}
 \end{subequations} 

where $a$ is a real constant, $H$ is an Hamiltonian, $c$ a lowering operator and $c^{\dagger}$ a raising operators. $F(H)$ is a functions of $H$ and can be in some applications a rational or even exponential function, one example is q-deformed algebras. Usually in applications in quantum physics $F(H)$ is a polynomial and then the algebra is called a polynomial Heisenberg-Weyl algebra. Some aspect of infinite and finite dimensional representations were discussed \cite{das91,que94} and constraints for the existence of finite dimensional unitary representations was provided. Those constraints take the form of systems of algebraic equations. Examples of representations which decompose into combination of finite and infinite representations were obtained \cite{and00,car04}. In applications, the generators $c$ and $c^{\dagger}$ take the form of higher order differential operators. Despite, the simplicity of such algebraic structure with three generators only and three defining commutator relations, it relates to large classes of isospectral and almost isospectral deformations of one dimensional Hamiltonian generalizing the harmonic and singular oscillators (including polynomial, and non polynomial deformations of the potential) \cite{car04, que14}. They include Painlev\'e transcendent (fourth and fifth) and generalisation \cite{nik97,car04, ber11, ber14,mar19}. It was demonstrated how this problem already at degree 6, the associated Chazy equation that defines the higher transcendental functions in which the Hamiltonian can be written is outside known Painlev\'e reduction \cite{mar19}. First degree operators and second degree operators lead to Lie algebra i.e. when $F(H)=f_0$ and $F(H)=f_0 +f_1 H$. The class of polynomial Heisenberg-Weyl algebras contains all generating spectrum algebras for quantum models related to exceptional Hermite and Laguerre polynomials \cite{que14} which then make the study of the related polynomial Heisenberg-Weyl algebra interesting from this perspective. Various other polynomial deformations of Lie algebras and their representations have attracted interest over the years \cite{van95, abd98,que07,isa14,mar20,cra21}.

Such structures share similarities with the Heisenberg-Weyl algebra of the harmonic oscillator that its generalize, which admit infinite dimensional representations. Polynomial Heisenberg-Weyl algebras allow for combinations of infinite and finite dimensional unitary representations. The representations is in general difficult to characterise and may depends upon solving systems of algebraic equations in order to find zero modes of lowering and raising generators, as both lowering and raising operator may admit several zero modes. Those lowering and raising operators are realized via higher order differential operators.

Hamiltonians possessing ladder operators with polynomial Heisenberg-Weyl algebras are also relevant in regard of constructing multi-dimensional Hamiltonians possessing integrals of motion, both integrable and superintegrable \cite{mar19,mar10,mar11}. Then progress on developing representations of the polynomial Heisenberg algebra can then have application in different context such as quantum integrable and superintegrable systems. We have recently demonstrated such algebraic structures give rise to complicated patterns of indecomposable representations taking the form of 2-chains when one consider certain Hamiltonians related to second order differential equations associated with exceptional Hermite polynomials and having ladder with polynomial Heisenberg algebra \cite{mar23}. This case corresponds to a particular case of rational solution for the model with the fourth Painlev\'e transcendent for specific values of the parameter $\alpha$ and $\beta$. This illustrate how rich is the representation theory for those.
 
This paper will deal with the case of induced representation from lowest states i.e. zero modes of the lowering operator and how the raising operator allows to generate additional states from the zero modes for a given Hamiltonian. Moreover, it will also deal with the case of highest weight for which the induced representations is achieved via a zero mode of the raising operator and then the action of the lowering operator provides the sequence of states from the point of view of the Hamiltonian. Here, the ladder operators, zero modes and Hamiltonian will involve the fourth Painlev\'e transcendent and then the induced representation will facilitate the construction of the states. We will establish several commutator identities involving monomials which have not been exploited in the literature. We will also present using this induced representations an explicit construction using differential operators. 

In Section 2, we present the general form of a polynomial Heisenberg algebra. In Section 3, we present induced construction and commutator identities. In Section 4, we apply the construction to the Hamiltonian with third order shape invariance in terms of fourth Painlev\'e transcedent. In Section 5, we present an explicit form of the induced construction in terms of polynomials written in terms of the fourth Painlev\'e transcendent and its derivative for lowest weight type and in Section 6 for the highest weight type.

\section{Polynomial Heisenberg algebra}

In many cases of Schr\"odinger equation, the potential is algebraic i.e. in terms of polynomials or rational functions and then explicit solutions for the Schr\"odinger equation

\begin{equation}
 H \psi_n = E_n \psi_n 
\end{equation}
can be associated with the theory of hypergeometric functions, and possibly generalizations in terms of confluent Heun or Heun equations via appropriate transformations. In some cases, when the potential depends on trigonometric or even elliptic functions an algebraic form for the potential may still be used via some algebraization transformation. However, even with an algebraic form, this is still a difficult problem in general to obtain exact solutions for the Schr\"odinger equation. This can be facilitated by using various algebraic methods and in particular the connection with ladder operators and algebraic structures such as the Heisenberg-Weyl type algebras may allow to get the complete spectrum from an initial state (zero modes) or many, as higher order operators can annihilate multiple states. The associated representations of the deformed Heisenberg-Weyl algebra can be constructed via the associated special functions and related recurrence relations. In later section, we will discuss how for Hamiltonians without an algebraic form and connecting with higher transcendental functions, one cannot rely on classical approaches to ODEs but still on an algebraic approach.
\newline
Another aspect of Heisenberg-Weyl algebra is the existence of Casimir invariant that can be exploited in regard of the descriptions of the representations. A Casimir invariant of the polynomial Heisenberg-Weyl algebra is a polynomial in the generators $\{c,c^{\dagger},H\}$

\begin{equation}
 K = K(c,c^{\dagger},H)= \sum_{i+j+k \leq n } \alpha_{ijk} H^i c^{j} (c^{\dagger})^k 
\end{equation} 
with $\alpha_{ijk}$ being constant such that
\begin{equation}
 [K,H]=0 ,\quad [K,c]=0 ,\quad  [K,c^{\dagger}]=0. 
\end{equation}
For the quadratic case $F(H)=b_2 H^2 + b_1 H + b_0$ the Casimir take the form

\begin{equation}
K= c c^{\dagger} - \frac{b_2}{3 a} H^3 - \frac{(b_1 +a b_2)}{2a} H^2 - \frac{(6b_0 +3 a b_1 + a^2 b_2)}{6a}H .
\end{equation}

It can be demonstrated that in fact the Casimir take the following form in general

\begin{equation}
 K= cc^{\dagger} - M(H) 
\end{equation} 
where the polynomial $M(H+a)$ is constrained by
\begin{equation}
 F(H)= M(H)-M(H-a). 
\end{equation}
 
This polynomial $M(H)$ can be constructed from the coefficient of the polynomial $F(H)$ of the commutator relations $[c,c^{\dagger}]$. As the Hamiltonian, the lowering and raising operators are differential operators this also implies that further relations in the realization can be obtained such as product ( alternatively to only commutator )
\begin{equation}
 c^{\dagger} c = M(H) ,\quad cc^{\dagger} = M(H+a).
\end{equation} 

This also imply in the differential operator realization the Casimir reduces to a polynomial of the Hamiltonian. Those additional relations in the realizations can be used to deduced the weight ( or energy from point of view of the corresponding Schrodinger equation ) of the zero modes. A zero mode is an eigenstate of the Hamiltonian such that the action of the raising or lowering ( or both ) is vanishing.

In this paper we will consider another approach, i.e. not building on Casimir, explicit realizations or factorization relations, but using induced representations constructions. This is an approach which is also based on establishing identities and commutator of monomials of the generators of the underlying quadratic algebra. Those formula will allow in later sections to provide further understanding of the explicit wavefunctions for the corresponding Hamiltonian and Schr\"odinger equation.

\section{Induced representations and algebraic definition of states}

The notion of induced representations has been widely studied in different context and in particular in regard of Lie algebras. However in regard of polynomial algebra much less is known. Constraints on the existence of zero modes of $c^{\dagger}$ and $c$ can be achieved via $cc^{\dagger}$ and $c^{\dagger}c$ and related polynomials in terms of the Hamiltonians. Here our approach differs and will concern lowest and highest weight constructions.

\subsection{Highest and lowest weight constructions}

We will define states via action on $\psi^{(i)}_{0}$ of the lowering operator $c$ and action of the Hamiltonian which plays a role analog of a Cartan generator with

\begin{equation}
 H \psi^{(i)}_0 = E^{(i)}_0 \psi^{(i)}_0 ,\quad c \psi^{(i)}_0 =0 .
\end{equation} 

Here $\psi_0^{(i)}$ ( with $i=1,...,l$ ) take into account the possibility of a set of $l$ zero modes, i.e. the set of states annihilated by $c$, which are the lowest states. Then we use the action of raising operators in following way

\begin{equation}
 \psi^{(i)}_{n} = (c^{\dagger})^n \psi^{(i)}_0 .
\end{equation} 

The construction of the induced representation consists in establishing the action of the generators on $\psi^{(i)}_{n}$

\begin{equation}
  c \psi^{(i)}_{n} = \alpha_{n} \psi^{(i)}_{n-1},\quad c^{\dagger} \psi^{(i)}_{n} = \psi^{(i)}_{n+1},\quad 
  H \psi^{(i)}_{n} = \beta_n \psi^{(i)}_{n} .
\end{equation}

The following identities can be demonstrated

\begin{equation}
 [H,(c^{\dagger})^k]= m(k) (c^{\dagger})^k  ,\quad [c,(c^{\dagger})^k]=  (c^{\dagger})^{k-1} R_k(H) .
\end{equation} 

We will provide further details in next subsection and determine explicitly the polynomial of H denoted $R_k(H)$. This polynomial depends on $k$. This allow In particular to demonstrate the following results

\begin{equation}
  c \psi^{(i)}_{n}  = [c,(c^{\dagger})^n] \psi^{(i)}_0 = R_n(E_0^i) \psi_{n-1}^{(i)},
\end{equation}
\begin{equation}
  H \psi^{(i)}_{n} = (c^{\dagger})^n H \psi^{(i)}_0 + [H,(c^{\dagger})^n] \psi^{(i)}_0 = ( E^{(i)}_0 + m(n) ) \psi_n^{(i)}.
\end{equation}
  
Using induced representations from highest weight of the form

\begin{equation}
 H \phi^{(i)}_0 = E^{(i)}_0 \phi^{(i)}_0 ,\quad c^{\dagger} \phi^{(i)}_0 =0 ,
\end{equation} 

and taking

\begin{equation}
  \phi^{(i)}_n= (c)^n \phi_0^{(i)},
\end{equation}  

the related action of generators is given by

\begin{equation}
  c \phi^{(i)}_{n} = \phi^{(i)}_{n+1},\quad 
 c^{\dagger} \phi^{(i)}_{n} = \tilde{\alpha}_n \phi^{(i)}_{n-1},\quad 
  H \phi^{(i)}_{n} = \tilde{\beta}_n \phi^{(i)}_{n}. 
\end{equation}

The results rely on establishing the following identities

\begin{equation}
 [H,c^k]= p(k) c^k ,\quad [c^{\dagger},c^k]= c^{k-1} S_k(H) .
 \end{equation}

 We will also establish in next subsection details on the polynomial of H denoted $S_k(H)$. This polynomial also depends explicitly on $k$. As a direct consequence one can obtain
 
\begin{equation}
c^{\dagger} \phi_n^{(i)} = c^{n-1} S_n(H) \phi_0^{(i)},
\end{equation}
\begin{equation}
H \phi_n^{(i)} = (E_0^{(i)} + p(n)) \phi_n^{(i)}.
\end{equation}    

One advantage to rely on induced representations in context of polynomial Heisenberg algebra is that for ladder operators which take the form of differential operators of degree 3 and higher, they are associated with higher transcendental functions. This means, in such cases, the Hamiltonian and ladder operators involve special functions only defined via nonlinear differential equations. Among them the well known Painlev\'e  transcendents and higher order analog. However, this means that action of ladder can no longer be straightforwardly calculated as for harmonic oscillator and its rational deformation and the zero mode are written in terms of those higher transcendental functions leading to solution only determined via iterative action of ladder operators and thus via the induced representations. However as commutator identities are determined the representation can be nevertheless determined explicitly. In Section 6, we will provide details on how those representations have a correspondence in terms of explicit polynomials of those higher transcendental functions and their derivative.

\subsection{Commutator identities for quadratic Heisenberg-Weyl algebra}

The purpose of this section is to consider the case of quadratic Heisenberg-Weyl algebra which connect with case of fourth Painlev\'e transcendent and related Hamiltonian and ladder operators of third order. We consider the following general quadratic Heisenberg-Weyl algebra formed by $\{H,c,c^{\dagger},1\}$

\begin{subequations}
\begin{align}
&  [H,c]=-ac \\
&  [H,c^{\dagger}]=a c^{\dagger} \\
&  [c,c^{\dagger}]= b_2 H^2 + b_1 H + b_0 .
\end{align} 
\end{subequations}

As consequence of the defining commutator relations, we obtain 

\begin{equation}
 [H,(c^{\dagger})^k]= 2k (c^{\dagger})^k 
\end{equation} 
\begin{equation}
 [H,c^k]=-2k (c)^k .
\end{equation} 

Those formula indicate analog of weight and that the Hamiltonian $H$ then play an analogous role as Cartan generator for the well-known Lie algebra $sl(2)$. Other commutator identities can also be demonstrated in different ways and in particular without relying on explicit differential operators realization or factorization relation. It can be shown that the commutator  of $c$ with monomial of $c^{\dagger}$ can be rewritten in following way
\begin{equation}
  [c,(c^{\dagger})^n]= (c^{\dagger})^{n-1} ( (a_0 + a_1 n ) H^2 + ( a_2 + a_3 n + a_4 n^2 ) H + ( a_5 + a_6 n + a_7 n^2 + a_8 n^3 ) = (c^{\dagger})^{n-1} R_n(H),
\end{equation}  

where the coefficient $a_0$ to $a_8$ are given by

\begin{equation}
 a_0 =0 ,\quad  a_1 = b_2 ,\quad a_2 =0\quad  , a_3= -a b_2 + b_1 ,\quad a_4 = a b_2 ,\quad a_5 =0,
\end{equation} 
\[ a_6 = \frac{1}{6} (a^2 b_2 + 6 b_0 - 3 a b_1),\quad a_7 = \frac{1}{2} (-a^2 b_2 +a b_1) ,\quad a_8 = \frac{a^2}{3} b_2. \]

We can also establish a similar formula for the commutator of $c^{\dagger}$ with a monomial of $c$

\begin{equation}
  [c^{\dagger},c^n]= (c)^{n-1} ( (a_0 + a_1 n ) H^2 + ( a_2 + a_3 n + a_4 n^2 ) H + ( a_5 + a_6 n + a_7 n^2 + a_8 n^3 ) = (c)^{n-1} S_n(H) ,
\end{equation}  

where the coefficient $a_0$ to $a_8$ take the form

\begin{equation}
 a_0 =0 ,\quad  a_1 =- b_2 ,\quad a_2 =0 ,\quad a_3= -a b_2 - b_1 ,\quad a_4 = a b_2 ,\quad a_5 =0, 
\end{equation}
\[ a_6 = \frac{1}{6} (-a^2 b_2 - 6 b_0 - 3 a b_1),\quad a_7 = \frac{1}{2} (a^2 b_2 +a b_1) ,\quad a_8 =- \frac{a^2}{3} b_2. \]

In the case of the Lie algebra $sl(2)$, the commutator identities of the triplet formed by the Cartan, the raising and the lowering generator play a role in construction of induced representations. This is then expected that polynomial identities will play a similar role for quadratic algebras.

\section{Representations Painleve IV: irreducible case $\beta>0$}

In this section, we recall some results related to third order shape invariance and construction based on second order supersymmetric quantum mechanics \cite{ber14,car04,mar10,mar11,and00,mar09}. The one-dimensional Hamiltonian with a third order ladder operators has the form (with $\lambda=1$, see \cite{and00,mar09}).

\begin{equation}
 H=-\frac{d^2}{dx^2} -2 f' + 4 f^2 +4 x f + x^2 - 1. 
\end{equation}

The functions $f=P_{IV}(x,\alpha,\beta)$ satisfy a second order nonlinear differential equation which can be written in terms of the fourth Painlev\'e equation. Here we will rely only on the functions f and its defining equation given by

\begin{equation}
 f''= \frac{f'^2}{2f} + 6 f^3 + 8 x f^2 + 2 (x^2 -(1+\alpha))f + \frac{\beta}{2f} . 
\end{equation}

The ladder operators of third order take the form

\begin{equation}
 c= M^{+} Q^{-} ,\quad  c^{\dagger}=Q^{+} M^{-} ,
\end{equation} 

where the operators $M^{\pm}$ and $Q^{\pm}$ are given by

\begin{equation}
M^{+}= \partial_x^2 + h(x) \partial_x +g(x) = (\partial_x + W_1)(\partial_x + W_2),
\end{equation}
\begin{equation}
M^{-}= \partial_x^2 - \partial_x h(x) + g(x) = (-\partial_x + W_2)(-\partial_x + W_1) ,
\end{equation}
\begin{equation}
 Q^{+}= (\partial + W_3),\quad  Q^{-}=(-\partial +W_3) .
\end{equation} 

and the functions $W_i$ for i=1,2,3 take the form 

\begin{equation}
 W_1=-f + \frac{f'-\sqrt{-\beta}}{2f} ,\quad W_2 =-f - \frac{f'-\sqrt{-\beta}}{2f} ,\quad W_3 = -2 f -x . 
\end{equation}  

The fact that the parameter takes positive or negative values is an important feature. In the case $\beta<0$ then the operators $M^{+}$ and $M^{-}$ well factorize into first order operators which allow in the context of supersymmetric quantum mechanics to be interpreted as physical intermediate Hamiltonians and in the case $\beta>$ the second order operators $M^{+}$ and $M^{-}$ would be referred as irreducible. The choice of sign for the parameter $d$ has also other consequence. Only in the case of $\beta<0$ the fourth Painlev\'e transcendent admit families of rational solutions known as Okamoto and generalised Hermite polynomials. Those special cases where studied in complete way in recent papers \cite{hus21,zel22}. The case of $\beta>0$ is of importance as it correspond to a physical model that do not reduce to algebraic form which is an interesting feature among the realm of exactly solvable quantum systems.

In order to establish explicit form for the states for the induced representation we will need to rely on further derivative of this nonlinear equations

\begin{equation}
 f^{(n)}= ( \frac{f'^2}{2f} + 6 f^3 + 8 x f^2 + 2 (x^2 -(1+\alpha))f + \frac{\beta}{2f} )^{(n-2)}. 
\end{equation}
 
Here $()^{(n)}$ is the nth derivative. We can then establish the following quadratic Heisenberg-Weyl algebra \cite{car04, ber14, mar10,mar11}

\begin{subequations}
\begin{align}
& [H,c]=-2 c \\
& [H,c^{\dagger}]=2 c^{\dagger} \\
& [c^{\dagger},c]= -2 (3 H^2 - (4\alpha +2) H + \alpha^2 +\beta) 
 \end{align}
 \end{subequations}

which correspond to the following choice of structure

\begin{equation}
 b_2=-6 ,\quad b_1=2(4\alpha+2) ,\quad b_2=-2 (\alpha^2 +\beta) 
\end{equation} 

where $\alpha$ and $d$ are parameter related to the fourth Painlev\'e transcendent.

Another framework in regard of higher order ladder was studied \cite{mar19}. This has lead to connection to Painlev\' via the Chazy equations and their reductions to Painlev\'e transcendents. 

\subsection{Lowest weight induced representations}

Considering the case where it is irreducible i.e. when the fourth Painlev\'e transcendent $P_{IV}(x,\alpha,\beta)$ does not admit any special solution in terms of rational or hypergeometric functions in the case $\beta>0$. In such case as the operators $M^{-}$ and $M^{+}$ do not have hermitian intermediate Hamiltonian the second order Darboux transformation can no longer be interpreted as two first order Darboux transformation. The factorized form can nevertheless be used to solve the equation related to the zero mode of the lowering operator. We will generate a infinite dimensional representations of lowest and highest weight type. The irreducible case considered occurs when $\beta$ is positive. Then one zero mode satisfying $c \psi_0^{(1)} =0$ and

\begin{equation}
 \psi_n= ( c^{\dagger})^n \psi_0^{(1)} 
\end{equation}

then the action of lowering and raising operator is given by

\begin{equation}
 c^{\dagger} \psi_n^{(1)}=\psi_{n+1}^{(1)} 
\end{equation}
\begin{equation}
c \psi_n^{(1)}=\left[ f(E_0^{(1)} \right] \psi_{n-1}^{(1)}.
\end{equation}

Following from general construction the commutator identities take the form

\begin{equation}
[c,(c^{\dagger})^n]=  (c^{\dagger})^{n-1} ( f_n(H) )
\end{equation}
\begin{equation}
 f_n(H)=   ( -2 \beta n -2n (2-2n+\alpha)^2 ) - 2 n(-8 +6n -4\alpha   ) H -6 H^2.
\end{equation} 

This leads to the explicit formula when acting on the zero mode

\begin{equation}
  [c,(c^{\dagger})^n]\psi_0^{(1)} = (c^{\dagger})^{n-1} ( f_n(E_0^{(1)}) ) \psi_0^{(0)} 
\end{equation}  

and in this case as $E_0^{(1)}=0$ the function $f_n(E_0^{(1)})$ can be written as

\begin{equation}
  f_n(E_0^{(1)})= -2 n(\beta + (2-2n +\alpha)^2 ) .
\end{equation}  

Those state formed by the lowest weight representation have interpretation has eigenstates of the corresponding Schrodinger equation as

\begin{equation}
H \psi_n =(E_0^{(1)} +2n) \psi_n .
\end{equation}

\subsection{Highest weight induced representations}

Then one zero mode satisfying $c^{\dagger} \phi_0^{(1)} =0$ and

\begin{equation}
 \phi_n= ( c)^n \phi_0^{(1)} 
\end{equation}

then the explicit action 

\begin{equation}
 c \phi_n^{(1)}=\phi_{n+1}^{(1)} 
\end{equation}
\begin{equation}
c^{\dagger} \phi_n^{(1)}=\left[ f(E_0^{(1)} \right] \psi_{n-1}^{(1)} .
\end{equation}

We then obtain the commutator identities

\begin{equation}
[c^{\dagger},(c)^n]=  (c)^{n-1} ( f_n(H) )
\end{equation}
\begin{equation}
 f_n(H)=   2n ( \beta + 4 (n-1)n + 4 (n-1) \alpha + \alpha^2 ) +2n( 4- 6 n -4 \alpha    ) H +6 n H^2.
\end{equation} 

When acting on the zero mode of the raising operator we obtain

\begin{equation}
  [c^{\dagger},(c)^n]\phi_0^{(1)} = (c)^{n-1} ( f_n(E_0^{(0)}) ) \phi_0^{(1)} .
\end{equation}  

As the energy of this zero mode is $E_0^{(1)}=\alpha-\sqrt{-\beta}$ we need for this case to consider irreducible cases among $\beta<0$. For this case

\begin{equation}
  f_n(E_0^{(1)})=  4n ( -\beta + 2 n^2 + n (-2 +3 \sqrt{-\beta}-\alpha )- \sqrt{-\beta} (2+\alpha) ).
\end{equation}  

This provides the construction of the induced representations in the highest weight case. The states formed by the highest weight representation have also an interpretation as eigenstates of the corresponding Schr\"odinger equation due to

\begin{equation}
H \phi_n =(E_0^{(1)} - 2n) \phi_n .
\end{equation}

\section{States for lowest weight representation as polynomials of fourth Painlev\'e transcendents}

In previous sections an algebraic description of the chain of states via induced representations was presented. One can also obtain explicitly the action of the different generators via various commutator identities. However, viewed as explicit expressions in terms of the function $f$ ( i.e. the fourth Painlev\'e transcendent ) it is a highly non trivial problem. This open problem has not been looked in the literature and only for different families of rational solutions and also hypergeometric type were studied and not the irreducible case. Looking at the irreducible case $\beta>0$ where

\begin{equation}
 \psi_0= e^{\int W_3(x') dx'}. 
\end{equation} 

We then obtain from the formula

\begin{equation}
 \psi_1^{(1)}= c^{\dagger} \psi_0^{(1)}
\end{equation} 

the following 

\begin{equation}
 \psi_1^{(1)}= \frac{1}{2 f^3} e^{\int W_3(x') dx'} \Big( 20 xf^5 + 8 f^6 + 4 f^4 (-3 +4 x^2 -3 f') + f' (\beta + f'^2) +f^3 (-8 x + 4 x^3 -8 x f'-2 f'')
\end{equation} 
\[ + f(\beta x + x f'^2 -2 f' f'') + f^2 (2 \beta +2 f'^2 -2 x f'' + f''') \Big). \]

Then using consequences of the defining second order nonlinear equation for the fourth Painlev\'e transcendent
\begin{equation}
 f^{(3)}= \left( \frac{f'^2}{2f} + 6 f^3 + 8 x f^2 + 2 (x^2 -(1+\alpha))f + \frac{\beta}{2f} \right)' 
\end{equation}
where $()'$ denotes $()_x$ and the equation of the fourth Painev\'e transcendent
\begin{equation}
 f^{(2)}=  \frac{f'^2}{2f} + 6 f^3 + 8 x f^2 + 2 (x^2 -(1+\alpha))f + \frac{\beta}{2f} 
\end{equation} 
we get an expression only in terms of f and f'. This allows to rewrite $\psi_1^{(1)}$ as

\begin{equation}
 \psi_1^{(1)} = \frac{e^{\int W_3(x') dx'} }{2f} \left(\beta-4 f(x+f)(-\alpha +f(x+f)+f'^2 \right). 
\end{equation} 

Here the monomials present are  $\{f^4, f^3, f^2, f, f'^2, 1\}$. For the next member of this sequence of states we consider $\psi_2$. Starting recursively with the formula

\begin{equation}
 f^{(6)}= \left( \frac{f'^2}{2f} + 6 f^3 + 8 x f^2 + 2 (x^2 -(1+\alpha))f + \frac{\beta}{2f} \right)'''',
\end{equation} 

and corresponding equations for $f^{(5)}$, $f^{(4)}$, $f^{(3)}$ and $f^{(2)}$, we get for $\psi_2^{(1)}$ an expression only in terms of $f$ and $f'$

\begin{equation}
 \psi_2^{(1)} = \frac{e^{\int W_3(x') dx'} }{f^2} \Big( 16 f^8 + 64 x f^7 + (96 x^2 -32 \alpha) f^6 +(32x +64 x^3 -96 x \alpha)f^5 - 8 f^4 f'^2 \end{equation}
\[ (-32 -8 \beta + 64x^2 +16 x^4 -96 x^2 \alpha +16 \alpha^2)f^4 -16 x f^3 f'^2\]
\[ + (-32x -16 \beta x+32 x^3 -32 x \alpha -32 x^3\alpha +32 x \alpha^2)f^3+ (-8x^2 +8\alpha) f^2 f' -16 f^2 f'\]
\[ + (-8 \beta -8 \beta x^2 +16 \alpha +8 \beta \alpha -32 x^2 \alpha -8 \alpha^2 +16 x^2 \alpha^2) f^2 +(-8 x + 8 x \alpha )f f'^2 +(-8 \beta x +8 \beta x \alpha )f + f'^4 +2 \beta f'^2 + \beta^2 \Big). \]

%\begin{figure}[ht]
%    \centering
%    \begin{subfigure}{0.31\textwidth}
%        \centering
%        \includegraphics[width=\linewidth]{fig3.jpg}
%        \caption{$\psi_3$}
%    \end{subfigure}
%    \hfill
 %       \begin{subfigure}{0.31\textwidth}
%        \centering
%        \includegraphics[width=\linewidth]{fig4.jpg}
%        \caption{ $\psi_4$}
%    \end{subfigure}
%    \hfill
%        \begin{subfigure}{0.31\textwidth}
%        \centering
%        \includegraphics[width=\linewidth]{fig5.jpg}
%        \caption{$\psi_5$}
%    \end{subfigure}

%\begin{figure}
%    \includegraphics[width=1.\textwidth]{fig3.eps}
%\end{figure}
%\begin{figure}
%    \includegraphics[width=1.\textwidth]{fig4.eps}
%\end{figure}
%\begin{figure}
%    \includegraphics[width=1.\textwidth]{fig5.eps}
%\end{figure}

\begin{figure}
\centering
\begin{minipage}{.28\linewidth}
  \includegraphics[width=\linewidth]{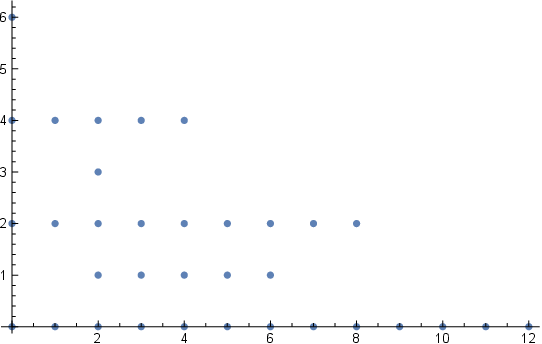}
\end{minipage}
\hspace{.05\linewidth}
\begin{minipage}{.28\linewidth}
  \includegraphics[width=\linewidth]{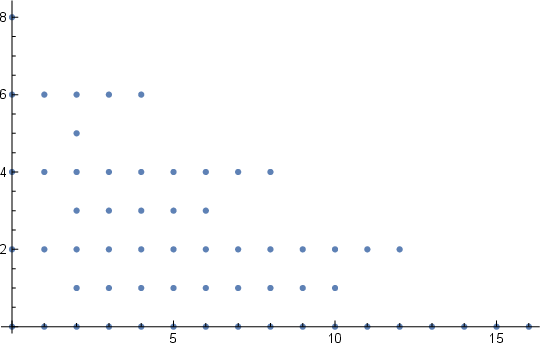}
\end{minipage}
\hspace{.05\linewidth}
\begin{minipage}{.28\linewidth}
  \includegraphics[width=\linewidth]{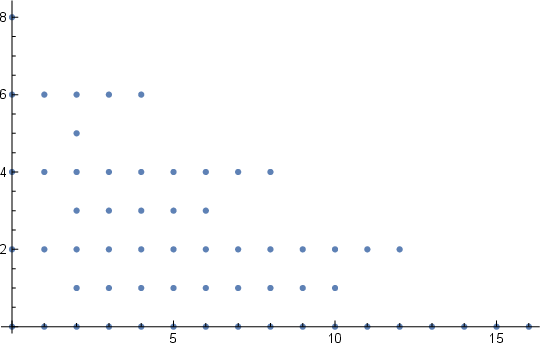}
\end{minipage}
\caption{Plot (in the x and y-axis) of the exponent $i$ and $2j$ (or $2j+1$) present for the expansion in terms of $f$ $f'$ for $Q_{1,ij}(f,f')$ and $Q_{2,ij}(f,f')$. The graph correspond to $\psi_2$, $\psi_3$ and $\psi_4$.}
\end{figure}

Considering the induced representation 

\begin{equation}
\psi_n^{(1)}= c^{n} \psi_0^{(1)}
\end{equation}

via explicit calculations from $\psi_2^{(1)}$, $\psi_3^{(1)}$, ..., $\psi_8^{(1)}$ we can obtain expansion in terms of $f$ and $f'$ only. However, if the expression can be determined explicitly the formula become quite large.  It was shown via software and symbolic calculations that for $\psi_n^{(1)}$ (up to $n=8$) that get polynomials in terms of f and f' with coefficient depending on x of the form

\[ \psi_{n}^{(1)}= \frac{e^{\int W_3(x') dx'}}{f^n} \left( \sum_{j=0}^{n} \sum_{i=0}^{4n-4j}  \alpha_{1,ij}(x) Q_{1,ij}(f,f') +  \sum_{j=0}^{n-1} \sum_{i=2}^{4n-6-4j}  \alpha_{2,ij}(x) Q_{2,ij}(f,f') \right) ,  \]
where
\[Q_{1,ij}(f,f')= f^{i} f'^{2j} ,\quad j=0,...,n; i=0,...,4n-4j,\]
\[ Q_{2,ij}(f,f')=f^{i} f'^{2j+1},\quad i=0,...,n-1; j=2,...4n-6-4j,\]

where $\alpha_{l,ij}(x)$ are polynomials in x for $l=1,2$ of degree at most $2n$. The problem of establishing the formula in general is quite complicated due to the growth of the number of terms and complexity.

\section{States for highest weight representation as polynomials of fourth Painlev\'e transcendents}

In previous sections an algebraic description of the chain of state via induced representation was presented with lowest weight. One can also obtain explicitly the action of the different generators via various commutator identities for the case of highest weight. However again viewed as explicit expression in terms of the function $f$ related to the fourth Painlev\'e transcendent it is highly non trivial. Due to the structure of the energy of the ground state we need to consider $\beta<0$

\begin{equation}
 \phi_0^{(1)}= e^{\int W_1(x') dx'} 
\end{equation} 
and then we obtain from
\begin{equation}
 \phi_1^{(1)}= c \phi_0^{(1)}
\end{equation} 

explicitly

\begin{equation}
 \phi_1^{(1)}= -\frac{e^{\int W_1(x') dx'} }{2f^3} \Big(8 x f^5 + 8 f^6 - 4 f^4 (2+3 f') + f' (\beta+f'^2) +f^3 (4 \sqrt{-\beta} x 
\end{equation} 
\[- 8 f' -2 f'') - 2 f f' f'' + f^2 (-2 \sqrt{-\beta} +2\beta +2 f' + 2 f'^2 + f''') \Big), \]

then using
\begin{equation}
 f^{(3)}= \left( \frac{f'^2}{2f} + 6 f^3 + 8 x f^2 + 2 (x^2 -(1+\alpha))f + \frac{\beta}{2f} \right)'
\end{equation}
\begin{equation}
 f^{(2)}=  \frac{f'^2}{2f} + 6 f^3 + 8 x f^2 + 2 (x^2 -(1+\alpha))f + \frac{\beta}{2f} 
\end{equation} 

we get polynomial only in f and f' and then $ \phi_1^{(1)}$ takes the form

\begin{equation}
 \phi_1^{(1)} =-\frac{e^{\int W_1(x') dx'} }{2f} \left( -2 \sqrt{-\beta} +\beta + 4 f ((1+\sqrt{-\beta})x -f (-1 +x^2 -\alpha + 2 x f +f^2)) +f'(2+f')\right). 
\end{equation} 
Considering $\phi_2^{(1)}$ we can obtain the following formula, again only in terms of $f$ and $f'$
\begin{equation}
 \phi_2^{(1)} = \frac{e^{\int W_1(x') dx'} }{f^2} \Big( -4 x f(x)^3 (-2 (\alpha  (\sqrt{-\beta}+2)+2 \sqrt{-\beta}+3)+2 (\sqrt{-\beta}+2) x^2+d +f'(x) (f'(x)+2))
\end{equation} 
 \[ +2 f(x)^2 (\beta (\alpha -3 x^2-1)+2 (-2 \alpha  (\sqrt{-\beta}+1)+(5 \sqrt{-\beta}+4) x^2 +\sqrt{-\beta})\]
\[ -(-\alpha +x^2-1) f'(x) (f'(x)+2))-2 f(x)^4 (-2 (\alpha  (\alpha +2)+\sqrt{-\beta}+2)\]
 \[+4 x^2 (\alpha +2 \sqrt{-\beta}+5)+ \beta +f'(x) (f'(x)+2)-2 x^4)+2 x f(x) ((\sqrt{-\beta}+2) f'(x) (f'(x)+2)-4 \sqrt{-\beta}+d (\sqrt{-\beta}+4))\]
 \[+\frac{1}{4} (\beta+f'(x)^2) (\beta-4 \sqrt{-\beta}+f'(x) (f'(x)+4))+8 x f(x)^5 (-2 \alpha -\sqrt{-\beta}+2 x^2-4)\]
 \[+8 f(x)^6 (-\alpha +3 x^2-1)+4 f(x)^8+16 x f(x)^7 \Big) .\]

\begin{figure}
\centering
\begin{minipage}{.28\linewidth}
  \includegraphics[width=\linewidth]{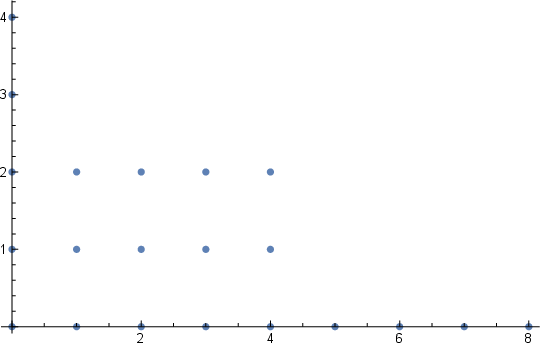}
\end{minipage}
\hspace{.05\linewidth}
\begin{minipage}{.28\linewidth}
  \includegraphics[width=\linewidth]{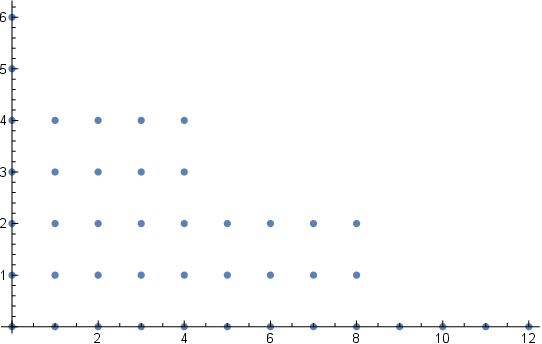}
\end{minipage}
\hspace{.05\linewidth}
\begin{minipage}{.28\linewidth}
  \includegraphics[width=\linewidth]{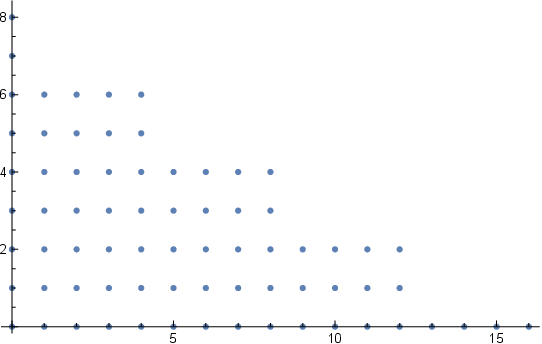}
\end{minipage}
\caption{Plot (in the x and y-axis) of the exponent $i$ and $2j$ (or $2j+1$) present for the expansion in terms of $f$ $f'$ for $Q_{1,ij}(f,f')$ and $Q_{2,ij}(f,f')$. The graph correspond to $\psi_2$, $\psi_3$ and $\psi_4$.}
\end{figure}

Considering the induced representation
\begin{equation}
\phi_n^{(1)}= c^{n} \phi_0^{(1)},
\end{equation}
it was shown via symbolic calculations that for $\psi_n^{(1)}$ (up to $n=8$) that we get polynomials in f and f' with coefficient depending on x of the form

\[ \phi_{n}^{(1)}= \frac{e^{\int W_1(x') dx'} }{f^n} \left(\sum_{j}^{n} \sum_{i=0}^{4n-4j}  \alpha_{1,ij}(x) Q_{1,ij}(f,f') +   \sum_{j}^{n-1} \sum_{i=0}^{4n-4-4j}  \alpha_{2,ij}(x) Q_{2,ij}(f,f') \right), \]
where
\[Q_{1,ij}(f,f')= f^{i} f'^{2j} ,\quad j=0,...,n; i=0,...,4n-4j,\]
\[ Q_{2,ij}(f,f')=f^i f'^{2j+1},\quad j=0,...,n-1; j=0,...4n-4-4j,\]

with $\alpha_{l,ij}(x)$ are polynomials in x for $l=1,2$ and of degree at most $2n$. The structure of those polynomials is also quite complicated. In that way, the induce representation can be used to obtain quite non-trivial solution from the point of view of the Schr\"odinger equation of the Hamiltonian and define polynomials in $f$ and $f'$ recursively via the action of the lowering or raising operators.
\newline
\newline
In view of results, the Hamiltonian in terms of the fourth Painlev\'e transcendent is exactly solvable, but the solution is not express in usual way.  They can be defined as solution being expressed in terms of orthogonal polynomials and more generally as solution of the hypergeometric equation (exact solvability) and Heun equations (quasi exact solvability). Here the solvability is provided via quadratic algebra and the possibility to define infinite dimensional representations. 

\section{N-dimensional superintegrable Hamiltonian with Painlev\'e transcendent and related states}

From the point of view of integrable and superintegrable systems, Hamiltonians involving higher transcendental functions are an important class. One can introduce a N-dimensional version based on taking a sum of one-dimensional Hamiltonians as building blocks in similar manner as for the isotropic harmonic oscillator or the Smorodinsky-Winternitz potential i.e. considering

\begin{equation}
 H=  \sum_i^N H_i = \sum_i^N  \left(-\frac{d^2}{dx_i^2} -2 f_i' + 4 f_i^2 +4 x_i f_i + x_i^2 -1 \right).
\end{equation}

Here all $\lambda_i=1$. The $\lambda_i$ can be exploited to provide anisotropic version. Here $f_i=P_{IV}(x_i,\alpha_i,\beta_i)$ which means it is not required for the construction to choose same parameters $\alpha_i$ and $\beta_i$ in different components, they can be taken independently. Integrals of motion related to separation of variables in Cartesian coordinates can be generated directly 

\begin{equation}
H_i =  -\frac{d^2}{dx_i^2} -2 f_i' + 4 f_i^2 +4 x_i f_i + x_i^2 -1 ,\quad i=1,...,N.
\end{equation}

This means the Hamiltonian is at least integrable as it allow separation of variables in Cartesian coordinates. For each component $H_i$ which depends on the variable $x_i$, the corresponding ladder are given by $c_i$ and $c_i^{\dagger}$. The ladder operators can be used to obtain integral of motion of the form 

\begin{equation}
I_{ij}= c_i c_j^{\dagger},\quad I_{ij}^{\dagger}= c_i^{\dagger} c_j,\quad i,j=1,...,N.
\end{equation}

This provide maximal superintegrability and existence of an underlying symmetry algebra. This Hamiltonian generalize the isotropic harmonic oscillator and the Smorodinsky-Winternitz systems in N dimensions. Here the construction of previous section can be used to construct the states in algebraic way (in the irreducible case) and provide explicit expression in terms of fourth Painlev\'e transcendent and derivative of the fourth Painlev\'e transcendent. The states relatives to a given components using lowest weight representations are denoted by

\begin{equation}
\psi_{i;n_i}^{1}= (c_i^{\dagger})^{n_i} \psi_{i,0}^{(1)}
\end{equation}

The formula in the case of lowest weight representation for all variables $x_i$'s

\begin{equation}
\phi_{n_1,...,n_N}= \prod_i^N (c_i^{\dagger})^{n_i} \phi_{0}^{[N]}
\end{equation}

where

\[ \psi_{0}^{[N]}= \prod_{i=1}^{N} \psi_{i;0}^{(1)}  \]

where

\[  c_i \psi_{i;0} =0 \]

and then by construction

\[  c_i \psi_{0}^{[N]} =0 ,\quad \forall i=1,...,N.  \]

The study of different integrable and superintegrable deformations is beyond the scope of this paper. However, this illustrates the wide applicability of the results from previous sections. So far only one example that would correspond to particular choice of $\alpha$ and $\beta$ in context of three-dimensional Euclidean space was solved algebraically \cite{mar22}. It was pointed out that the symmetry algebra is a generalization of the $su(3)$ algebra and the finite dimensional representations takes the form of multiplets that can be decomposed into $su(3)$ like multiplets. However, as the parameters of the fourth Painlev\'e transcendent were such that rational solutions exist this allowed to make calculation using explicit realizations for the wavefunctions to build the representations. When the parameters are generic as pointed out in earlier sections induced representations need to be used and the explicit wavefunction take a complicated form.

\section{Conclusion}

Polynomial Heisenberg-Weyl algebras and their constraints for the existence of finite dimensional representations  have been studied. Those are important in applications to quantum mechanical systems as they correspond often to degenerate spectrum and their related states decomposed into multiplets. Here we provided insight into induced construction of infinite dimensional representations and their related coefficients for action of the generators. In order to establish explicit formula We have used identities based on commutators of monomials of the generators and applied on the Hamiltonian related to the fourth Painlev\'e transcendent. This Hamiltonian possesses third order ladder operators and also connect with certain deformations of harmonic oscillator. The representations of highest and lowest weight types become in this setting two variables polynomials in terms of the fourth Painlev\'e transcendent and its derivative. To our knowledge those types of polynomials have not been studied in the literature. They may have broader applications in similar way as other type of polynomials ( among them Lam\'e polynomials in terms of elliptic functions ) appear in different context of mathematical physics.

\section*{Acknowledgement}
IM was supported by by Australian Research Council Future Fellowship FT180100099.

\end{document}